\documentclass[10pt, a4paper]{article}

\usepackage[final]{lrec2026}
\usepackage{amsmath}
\usepackage{booktabs}
\usepackage{tabularx}
\usepackage{import}
\usepackage{xcolor}
\usepackage{graphicx}

\title{S2Doc - Spatial-Semantic Document Format}

\name{Sebastian Kempf, Frank Puppe}

\address{University of Würzburg \\
Center for Artificial Intelligence and Data Science (CAIDAS)\\
Emil-Fischer-Straße 50\\
97074 Würzburg, Germany \\
         \{sebastian.kempf, frank.puppe\}@informatik.uni-wuerzburg.de\\
}

\abstract{
Documents are a common way to store and share information, with tables being an important part of many documents.
However, there is no real common understanding of how to model documents and tables in particular.
Because of this lack of standardization, most scientific approaches have their own way of modeling documents and tables, leading to a variety of different data structures and formats that are not directly compatible.
Furthermore, most data models focus on either the spatial or the semantic structure of a document, neglecting the other aspect.
To address this, we developed \textit{S2Doc}, a flexible data structure for modeling documents and tables that combines both spatial and semantic information in a single format.
It is designed to be easily extendable to new tasks and supports most modeling approaches for documents and tables, including multi-page documents.
To the best of our knowledge, it is the first approach of its kind to combine all these aspects in a single format.
\newline  }
\begin{document}

\maketitleabstract

\section{Introduction}
Extracting and understanding documents has always been a common task in the domain of artificial intelligence.
From OCR and text recognition to document layout analysis and information extraction, many different tasks can be performed on documents.
With the rise of large language models, interest in document understanding has increased, since these models benefit greatly from additional context and structure provided by documents.
Tables are of particular interest because they contain a large amount of structured information.
However, tables are designed for human consumption, so automatic processing and understanding of tables is challenging.

While tables and documents are subject to active research, there is no real common understanding of how to model documents and tables in particular.
Usually, every approach has its own way of modeling the documents and tables, leading to a variety of different data structures and formats that are not compatible with each other.
While some areas have established some kind of standard, like the \textit{COCO} format for image segmentation, these are task-specific and cannot be easily adapted to other tasks.
This lack of standardization causes several problems.
Implementing different approaches, or combining them, is often difficult or impossible due to incompatible data structures or missing information.
Creating datasets is also more challenging because there is no clear consensus on how to model a table or document.
As a result, datasets are often limited to a specific task, so evaluating the performance of whole pipelines is difficult.
In addition to these problems, most data models and datasets focus on either the spatial or the semantic structure of a document.
However, both are important for a complete understanding of a document and should therefore be jointly considered, modeled, and evaluated.
Furthermore, one often overlooked aspect is the multi-page nature of most real-world documents.
For example, scientific articles often contain dozens of pages, with tables and figures spread across multiple pages, so a complete understanding of the document requires considering all pages.
Most existing data structures focus on individual pages and neglect the overall document context.

Because of that, we developed \textit{S2Doc}\footnote{\url{https://github.com/Raynswor/s2doc}}.
It is a modular and flexible data structure for modeling documents and especially tables, combining both spatial and semantic information in a single format.
It has been designed to be easily extendable to new tasks and comes with support for most of the modeling approaches of documents and tables as well as multi-page documents.
It tries to establish a common ground for modeling documents and tables to make implementing, combining, and evaluating document and table understanding approaches easier.
To the best of our knowledge, it is the first approach of its kind.

\section{Related Work}
As previously mentioned, there is no real standard for modeling documents and tables in the scientific community.
Most data structures are means to an end to solve a specific task and are implemented to fit these requirements.
Because it is not feasible to cover all existing approaches, we will focus on existing document and table formats, focusing on their advantages and disadvantages with regards to the mentioned problems.
A tabular comparison of the different approaches can be found in Table~\ref{tbl:comparison}.

The categorization of existing document and table formats is not trivial, as there are many different ways to categorize them.
Because the focus of this work is on practicality and usability, we will loosely categorize them based on their main use case and application.
We identified the following categories:
(1) OCR-centric, (2) human-centric, (3) data-centric, (4) table-centric, and (5) other.

\subsection{OCR-centric}
These formats are primarily designed to represent the results of digitalization pipelines, including document types such as scans, images, and especially OCR output.
Their focus is capturing the physical layout and textual content of documents and uses hierarchies of elements to represent the document structure.
They are machine-readable and can be easily processed by algorithms, but focus on layout rather than semantic information.
Prominent examples of this category are \textit{PAGE}, \textit{ALTO}, and \textit{hOCR}.

\textit{PAGE} by \cite{pletschacher_page_2010} is a popular XML standard to encode digitised documents in the domain of document image analysis and OCR.
It is designed to support the representation and evaluation of intermediate results of these tasks.
In addition to the layout structure and page content, it also provides the possibility to encode image characteristics and processing steps.
It has been used in competitions like the ICDAR Page Segmentation.

Similarly, \textit{ALTO} by \cite{alto} is another XML Schema that focuses on layout preservation and content representation.
Additionally, it captures structural elements such as margins, headings, and other layout features.
It is widely used in libraries and archives for digitized documents, especially in conjunction with the METS format for storing and exchanging digital objects.

\textit{hOCR} by \cite{breuel_hocr_2007} is an HTML microformat to represent intermediate and final OCR results.
It offers several markups, allowing for modeling the logical structure of documents as well as typographical and layout information.
Because it is based on HTML, it can be easily visualized and processed with corresponding tools.

\subsection{Human-centric}
This category includes formats that are primarily designed for human consumption.
They focus on the visual appearance of documents and sometimes the ability to edit them.
Usually, they are not easily machine-readable and do not provide a clear structure for the document content (with the exception of HTML).
They emphasize logical structure and rich text with control over layout.

\textit{DocX} and \textit{ODT} are widely used formats for word processing documents based on XML.
They are designed to be easily editable and provide multiple formatting options.
However, any formatting information is only represented visually and not accessible in a structured way, making automatic analysis of these documents difficult and their use in document understanding pipelines limited.

Documents in these formats are often exported as \textit{PDF} documents for sharing and publishing, because \textit{PDF} preserves the visual appearance across platforms and devices.
However, \textit{PDF} is exclusively focused on the visual representation of documents and does not provide a clear structure for the document content.
It is a binary format encoding document elements as a sequence of drawing commands, so even words are not explicitly represented, but have to be inferred from the layout.
This makes automatic analysis of \textit{PDF} documents very challenging.

\textit{HTML} is a markup language, primarily used for creating web documents and applications.
A document is seen as a tree of elements, where each element can have attributes and nested elements.
It provides a wide range of tags to represent different types of content, such as headings, tables, and lists.
It is also used in other contexts than the world wide web, for example, as a data format to structure tables in extraction pipelines.
Because of its structure and tagging system, it is easily machine-readable and can be processed by algorithms.

Another markup language is \textit{\LaTeX}, which is widely used in academia for writing scientific documents.
Rather than modeling the document visually, it focuses on describing the logical structure and content.
Because it contains a special environment to represent tables, the logical structure of them as well as the textual content can be extracted relatively easily.
Beyond that, the physical and functional models of a table are not represented at all or available in a standardized way.

\subsection{Table-centric}
In contrast to the previous categories, this category of formats focuses exclusively on tables and tabular data.
This includes both formats to design and represent tables as well as formats to store and exchange tabular data.
However, the origin of these formats lies in the tabular nature of data rather than the tables themselves.

Spreadsheet formats like \textit{XLSX} and \textit{ODS} are widely used for creating and editing tables.
They provide a grid-like structure to organize data in rows and columns, along with various formatting options.
However, like their word processing counterparts, they focus on visual representation and editing capabilities rather than structured data representation.
Furthermore, it is up to the user to define what a table is and how it is structured, as the format does not enforce any specific structure.

Simpler versions of these formats are \textit{CSV} and \textit{TSV}, which are plain text formats to store tabular data and use defined characters to separate values and rows.
They are easy to read and write, making them suitable for data exchange.
However, they do not provide any information about the structure or meaning of the data, making them less suitable for complex tables.

\subsection{Data-centric}
Data-centric formats are designed for storage and exchange.
Their focus is to represent structured data, primarily to provide it in a machine-readable way.

\textit{JSON} and \textit{XML} are two widely used formats for representing structured data, the latter having been mentioned before in the context of OCR-centric formats.
They both allow for representing hierarchical data structures and are adjustable to different use cases, making them suitable to represent tables as well as documents.
However, they do not provide any specific structure or semantics for tables or documents, so it is up to the user to define how to represent them.

In the context of computer vision and image understanding, the \textit{COCO} format by \cite{lin_microsoft_2015} has become a popular standard for representing annotated images.
It is primarily designed for object detection and segmentation tasks, representing the spatial location of objects within an image using bounding boxes and segmentation masks in JSON format.
It has also been adapted to represent document layouts, table detection, and recognition.
No semantic or structural information can be modeled beyond the object classes, making it unsuitable for tasks beyond the previously mentioned.

\subsection{Other}
A whole different category of data-centric 'formats' are knowledge graphs and ontologies.
In contrast to all previously mentioned formats, they are not designed to represent documents or tables, but to represent knowledge using entities and their relationships.
While they are not directly applicable to document and table understanding tasks, they can be used to represent the semantic information contained in documents and tables.
\textit{RDF} and \textit{OWL} are two standard ways to represent knowledge graphs and ontologies.

Not necessarily a format, but nonetheless belonging to this category are \textit{relational databases}, which store data in tables with defined schemas.
Consequently, they are not designed to represent tables visually, but rather to manage the underlying data.
While they are ideal for storing and querying structured data, they are not suitable for representing complex tables with hierarchical structures without additional processing.

To summarize, most existing formats focus on the spatial and logical aspects of documents and tables.
Because they are usually designed for specific tasks and use cases, they are inflexible and difficult to adapt to new requirements.
Modeling semantic information is often not considered, which means that an important part of the data is lost.
On the other hand, formats that do consider semantic information cannot represent the spatial structure of documents and tables.
This situation is further complicated by the fact that many real-world documents are multi-page, which is often not supported by existing formats.
Because of that, the need for a format that combines everything is evident.

\begin{table}[!htb]
  \centering
  \scriptsize
  \setlength{\tabcolsep}{3pt}
  \begin{tabularx}{\columnwidth}{l*{7}{>{\centering\arraybackslash}X}}
    \toprule
    \textbf{Format} & \textbf{Phys.} & \textbf{Log.} & \textbf{Sem.} & \textbf{Cont.} & \textbf{Read.} & \textbf{Fidelity} \\
    \midrule
    PAGE XML & Yes & Yes & Lim. & Part. & Yes & H \\
    ALTO XML & Yes & Yes & Lim. & Part. & Yes & H \\
    hOCR & Yes & Yes & Lim. & No & Yes & H \\
    \midrule
    DOCX/ODT & Flow & Yes & Lim. & Yes & Part. & M \\
    PDF & Yes & Part. & Part. & No & No & H \\
    HTML & No & Yes & Lim. & Yes & Yes & M \\
    LaTeX & No & Yes & No & Yes & Yes & L \\
    \midrule
    XLSX/ODS & Flow & Yes & Lim. & Part. & Part. & M \\
    CSV/TSV & No & Lim. & No & No & Yes & - \\ 
    XML/JSON & No & Sch. & Sch. & Sch. & Yes & - \\
    \midrule
    COCO & Yes & No & Lim. & No & Yes & L \\
    RDF/OWL & No & No & Yes & No & Yes & - \\
    Relational DB & No & Yes & Yes & No & Yes & - \\
    \bottomrule
  \end{tabularx}
  \caption{Comparison of different document and table data structures (s. Section~\ref{sec:s2doc}).
  \textbf{Phys.}: Physical Structure.
    \textbf{Log.}: Logical Structure.
    \textbf{Sem.}: Semantic Structure.
    \textbf{Cont.}: Table context.
    \textbf{Read.}: Machine Readability.
    \textbf{Fidelity}: Layout Fidelity.
    Lim.: Limited, Sch.: Schema-based, Part.: Partial, Flow: Flow-based
  }
  \label{tbl:comparison}
\end{table}

\section{S2Doc} \label{sec:s2doc}

\textit{S2Doc} is primarily inspired by the works of \cite{wang_tabular_1996,hurst_interpretation_2000}, who developed an abstract model for tables.
They identified four main models of a table:
(1) physical, (2) logical, (3) functional, and (4) semantic.
The physical model describes a table as a set of cells, with each cell having a spatial position within a page.
The logical model arranges the cells into rows and columns, forming a grid-like structure.
The functional model defines the purpose of each cell, meaning if it is a label or a data cell.
The semantic model, called abstract table by \cite{wang_tabular_1996}, combines the previous models to build a representation of the table that is 
\clearpage
\begin{figure*}[]
  \centering
  \includegraphics[width=\textwidth]{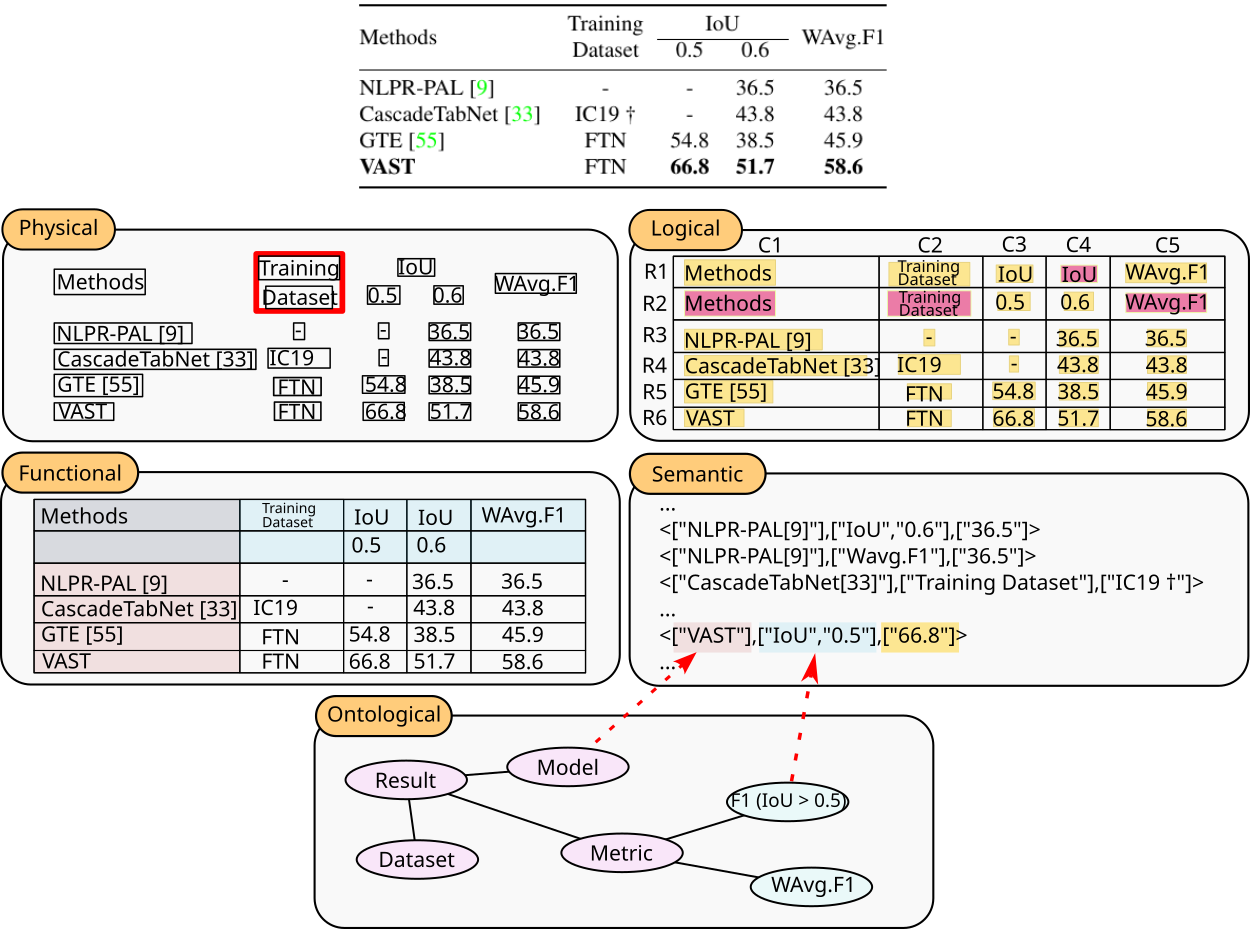}
  \caption{Example table from \cite[p. 7]{huang2023improvingtablestructurerecognition} and its five models.
  The physical model is a set of bounding boxes that may have content.
  Because it is assumed that the input medium is a PDF, the cell content is known.
  Depending on the extraction approach, 'Training' and 'Dataset' may be represented as two separate cells or as one cell.
  The logical model orders these cells into rows and columns, either using a grid/matrix structure or a graph structure.
  The graph model connects cells in the same row or column, either directly or indirectly using intermediate nodes/objects.
  Depending on the approach used for structure recognition, there may be intermediate states of logical structure.
  For example, it is not unambiguous, which rows 'Methods' and 'WAvg.F1' are in, because they are in between two of them.
  They could be put in one of the two columns, put in between the two rows, or be marked as a spanning cell.
  The same applies to the 'IoU' cell, which is between two columns.
  The functional model identifies row and column label cells based on the results of the previous models.
  The semantic model combines all previous models to form a set of tuples $\langle [ \textrm{row header(s)} ], [ \textrm{column header(s)} ], [ \textrm{value} ] \rangle$.
  Within this step, the resolution of ambiguities from the logical model is crucial, as it directly affects the resulting tuples.
  If 'IoU' was not identified as belonging to both 'C3' and 'C5' columns, instead of $\langle [ \textrm{VAST} ], [ \textrm{IoU},\textrm{0.5} ], [ \textrm{66.8} ] \rangle$, 'IoU' would be missing.
  Finally, the ontological model is the annotation of the table with background knowledge.
  The background knowledge is not contained in the table itself, but may be provided by the surrounding document or an external knowledge base.
  The annotation is achieved by associating elements with concepts and entities.
  In this example, 'VAST' is associated with the concept 'Model' and 'IoU' with the entity 'IoU'.
  }
  \label{fig:example}
\end{figure*}
\clearpage
\noindent
independent of its visual appearance.

While this model is suitable for modeling tables, it is not enough to get a real understanding of a table.
The semantic model of a table is supposed to capture the meaning of a table, but it is limited to the information contained in the table itself.
Although tables should be self-explanatory (which they often enough are not), in order to be fully understood, they require additional context which may either be provided by the surrounding document or an external knowledge base.
In other words, there has to be some kind of ontological annotation of the elements to come closer to a real understanding.
Using the example in Figure~\ref{fig:example} from \cite[p.7]{huang2023improvingtablestructurerecognition}, the string "WAvg.F1" is not explained anywhere in the table.
In this case, this string is explained one page before as the "Weighted Average F1 Score".
Albeit a very simple example, it illustrates two things:
the need for explicit ontological modeling, and
the need for contextualization, especially of tabular elements.

To this end, the idea of modeling tables using different structures has been adapted and extended to model entire documents, with further inspiration drawn from other formats like \textit{PAGE} and \textit{HTML}.
It is a modular and flexible data structure for modeling documents and especially tables, combining spatial, logical, and semantic information in a single format.
Designed to be easily extendable to new tasks, it is suitable for a wide range of document and table understanding applications and tries to establish a common ground for modeling documents and tables to make implementation and evaluation easier.

\subsection{Main Idea}
The main idea of \textit{S2Doc} is to represent a \texttt{Document} as a collection of \texttt{Page} objects, where each contains a set of \texttt{Element} objects, which may reference each other and may be annotated with semantic information.
A \texttt{Page} describes a section of a document, for example, a page in a PDF, an image, or a slide in a presentation.
\texttt{Element} describes a document entity, such as a token, a table cell, a paragraph, or an image.
Each element has a unique, document-specific identifier that can be used to reference it from other elements or even documents.
Beyond its associated region that describes its location, an element has a type, defining what kind of entity it represents, and a data dictionary to store arbitrary additional information.
Neither the type nor the data dictionary of an element is restricted in any way or enforced to follow a specific schema by \textit{S2Doc}, enabling easy adaptation to new tasks and requirements.
It should be noted that while document entities could be represented using this general class, \texttt{Element} should primarily serve as a superclass to be inherited from.
For example, a \texttt{Table} can inherit from \texttt{Element} and thereby represent a table as a specialized element with additional attributes and methods like the number of rows and columns or methods to access specific cells.
Understanding a given adaptation of \textit{S2Doc} therefore only requires knowledge about the specialized classes and their attributes, while the general structure of the document remains the same.

\subsection{Physical/Spatial Structure}
Each \texttt{Page} is described by at least one \texttt{Space}, defining dimensionality and size of a coordinate system.
This is important because it enables the representation of documents in different coordinate systems at the same time.
For example, usually a PDF uses a coordinate system that is based on points and not pixels that are the standard unit of measurement for digital images.
Rendering a PDF page as an image at a certain resolution changes the coordinate system to a pixel-based one.
If both representations are of interest, two \texttt{Spaces} can be defined for one \texttt{Page}.
It is therefore possible to localize elements extracted directly from the PDF, as well as elements that have been extracted from an image rendering of the same page.
In order to localize elements within a \texttt{Space}, it is necessary to define their \texttt{Region}, a geometric shape demarcating a section of \texttt{Space}.
Beyond the common geometric shapes like rectangles and polylines, it is also possible to define 1-dimensional regions in a suitable space.
This allows, for example, to represent text spans in a text-based document, therefore enabling the representation of text documents without visual layout.

The top left box in Figure~\ref{fig:example} shows the physical model of the example table above.
Each black box represents a \texttt{Region} of an \texttt{Element} (here \texttt{TableCells}) that may contain some kind of content.
In this case, the input medium is a PDF document, so the content of each cell is known.
The example shows a disambiguity in the physical model, as it is not clear if 'Training' and 'Dataset' are two separate cells or one cell spanning two rows (marked in red).
Some approaches may represent them as one cell, while others may represent them as two separate cells, but both can be modeled.

\subsection{Logical Structure}
Documents usually follow a hierarchical structure.
For example, a document contains pages, containing paragraphs, which contain sentences, containing words, and so on.
\textit{HTML} leverages this hierarchical structure by representing a document as a tree of elements, where each element can have attributes and nested elements.
However, this structure is neither suitable for all document types and tasks, nor for tabular modeling.
For example, a table cell lies in a row and a column at the same time.
This is not representable in a tree structure where nodes can have only one parent.
This is also the reason why \textit{HTML} has an exclusively row-based representation of tables.

\textit{S2Doc} follows a different approach by representing the logical structure of a document as a directed acyclic graph (DAG) called \texttt{ReferenceGraph}.
By design, a \texttt{Document} would be a virtual 'root' node, with \texttt{Pages} as its children.
Beyond that, there are no restrictions on how to structure the document logically.
A string could be part of a table cell belonging to a table on the same page, while this table is associated with two pages, because it spans multiple pages.
This allows to represent any kind of logical structure, including hierarchical and non-hierarchical relationships between elements.
While this mechanism is intended to represent the logical structure of a document as defined by 'belongs to' relationships, it could also be used to represent other kinds of relationships between elements, such as the reading order of text elements.

Applying this to the running example, the box denoted as \textit{Logical} in Figure~\ref{fig:example} shows two possibilities of modeling a logical structure of the example table.
It is assumed that a \texttt{Table} element has been created for the table itself and associated with the corresponding \texttt{Page}.
The first possibility is to represent the table as a grid/matrix structure, where each cell is associated with its corresponding row and column.
The second possibility is to represent the table as a graph structure, using \texttt{Row} and \texttt{Column} elements to represent rows and columns and associating the cells with them.
In this case, column \texttt{C5} would be associated with the cells with content \texttt{WAvg.F1,26.5,43.8,45.9,58.6}.

The example table also contains logical ambiguities because it is not clear which rows 'Methods' and 'WAvg.F1' belong to, and which column 'IoU' belongs to.
All three are in between the two rows/columns that are formed by the remainder (marked in red).
Consequently, depending on the approach used for structure recognition, there may be intermediate states of logical structure.
For example, 'IoU' could be put in one of the two columns, or be put in between them.
However, the final logical structure should be unambiguous, as it is the basis for the subsequent functional and semantic models, so every element should appear in every row and column it belongs to.
In this case, 'IoU' should be both columns \texttt{C3} and \texttt{C4}, while 'Methods' and 'WAvg.F1' should both respectively appear in rows \texttt{R1} and \texttt{R2}.
Because situations like these can almost always only be resolved using some kind of visual indicator or contextual knowledge, the ability to model everything in a single format is even more important.

\subsection{Functional Structure}
The functional model defines the purpose of each element within a document.
For tables, this means identifying header and data cells according to the original table model.
However, this is not enough to later extract the semantic model of a table, especially when dealing with complex tables.
With \textit{S2Doc} and the attribute dictionary of \texttt{Elements}, it is possible to store arbitrary additional information about an element.
So beyond the simple classification of header and data cells, there may be more fine-grained classification that is relevant for the task at hand.

So in the running example (Figure~\ref{fig:example}, 'Functional'), there has been made a distinction between row and column header cells.
In practice, this would mean that the corresponding \texttt{Elements} (be it \texttt{Row} or \texttt{TableCell}) have an attribute like \texttt{isRowHeader} or \texttt{isColumnHeader} set to true.

\subsection{Semantic Structure}
The semantic structure is the culmination of all previous models, combining them to form a representation of the document that is independent of its visual appearance.
For tables, this means extracting the information contained in the table and representing it as a set of tuples $\langle [ \textrm{row header(s)} ], [ \textrm{column header(s)} ], [ \textrm{value} ] \rangle$.
Every tuple entry is a list, because there may be multiple row and column headers associated with a value.
While there is usually only one value per cell, there may be situations where a cell contains multiple values, for example if table cells are modeled in an unusual way.
Depending on the given model, this structure can be computed for tables by either traversing the references in the \texttt{ReferenceGraph} or by using a matrix-like structure attached to the \texttt{Table} element.

In the running example (Figure~\ref{fig:example}, 'Semantic'), the semantic model of the example table is shown as a set of tuples.
This example illustrates the importance of resolving ambiguities in the logical model, as it directly affects the resulting tuples.
If 'IoU' was identified as only belonging to \texttt{C4} and not \texttt{C3}, the tuple $\langle [ \textrm{VAST} ], [ \textrm{IoU},\textrm{0.5} ], [ \textrm{66.8} ] \rangle$ would be missing 'IoU' in the column headers.

However, it is also possible to use the semantic model to check the consistency of the logical model.
Because the semantic model is independent of the visual appearance of the table, similar tables in the same document or across documents should have a similar semantic model.
For example, if another table in the same document has column tuple entries $\langle [ \textrm{IoU},\textrm{0.5} ] ] \rangle$ and $\langle [ \textrm{IoU},\textrm{0.6} ] \rangle$,
the possible mistake of associating 'IoU' with only one column in our example could be identified.

\subsection{Ontological Structure}
As mentioned above, real understanding needs contextualization, either provided by the surrounding document or an external knowledge base.
Using the previous structures, the table and surrounding document can be modeled, but the actual meaning of the elements is not yet represented.
In order to represent this connection between elements and their meaning, \textit{S2Doc} adds another layer using the concept of a \texttt{SemanticKnowledgeGraph}.

A \texttt{SemanticKnowledgeGraph} is a graph structure, representing knowledge using \texttt{SemanticConcept}, \texttt{SemanticEntity} and \texttt{SemanticRelationships} to connect them.
It is independent from the document structure and can be used to represent any kind of knowledge from any kind of origin.
A \texttt{SemanticConcept} represents an abstract idea or category, like 'Person' and is the equivalent of a class in an ontology modeled in \textit{RDF} or \textit{OWL}.
A \texttt{SemanticEntity} represents a specific instance of a concept, like 'Albert Einstein', and is the equivalent of an individual in an ontology.
Concepts and entities form a graph structure with \texttt{SemanticRelationships} serving as edges.
For example, if one has a set of documents containing tables with identical headers but different values, one could create the semantic graph for one of them and then reuse it for others.
One other idea from ontologies and knowledge graphs has been adopted, namely the use of \texttt{Literals} to represent primitive values like strings, numbers, and dates as entities in the graph and thereby enabling the representation of relationships between these values and other entities if necessary.

The connection between the document structure and this new ontological structure is achieved by another \texttt{ReferenceGraph}, called \texttt{SemanticReferenceGraph}.
It connects \texttt{Elements} from the document structure with \texttt{SemanticConcepts} and \texttt{SemanticEntities} from the ontological structure, thereby associating elements with their meaning.
This is again not restricted, so elements can be associated with multiple concepts and entities.
Elements can be associated with concepts to represent uncertainty and make inference more flexible.
For example, if a table cell contains the value 'Apple', it could refer to the fruit or the company, so both entities are associated with the cell.
Assuming that this column is associated with the concept 'Company', one could infer that the cell refers to the company and not the fruit.

The example in Figure~\ref{fig:example} (box denoted as \textit{Ontological}) shows a possible ontology that could have been extracted from the document or given as an external knowledge base.
A concept 'Result' has been associated with three other concepts called 'Model', 'Dataset', and 'Metric', the latter of which has been associated with two entities 'IoU' and 'WAvg.F1'.
The actual annotation to elements is shown as a red dashed line.
In this case, the cell with content 'VAST' has been associated with the concept 'Model', while the cell with content 'IoU' has been associated with the entity 'IoU'.
It is important to note that this annotation is possible across the different models of document structure.

\section{Workflow Example}

The modular and flexible structure of \textit{S2Doc} allows modeling the entire workflow of a document and table understanding pipeline.
This includes the initial extraction of elements from the input medium, the analysis of the document structure, the extraction of the different table models, and the semantic annotation of elements.
All pipeline steps are explicitly represented by the different models, making the transformation more generic, validatable, and reproducible.
Furthermore, it is possible to represent uncertainty and knowledge at each step of the process because each element can be annotated with confidence scores and associated with multiple concepts and entities if necessary.
This is especially important for table transcription tasks, where the input medium may not provide all necessary information, for example, if the input is a scanned document or an image.
However, even if the input is a digital document such as a PDF or HTML, there are ambiguities and uncertainties that need to be resolved, which is why the ability to represent everything in a single format is crucial.

\begin{figure}[!htb]
  \centering
  \includegraphics[width=0.83\columnwidth]{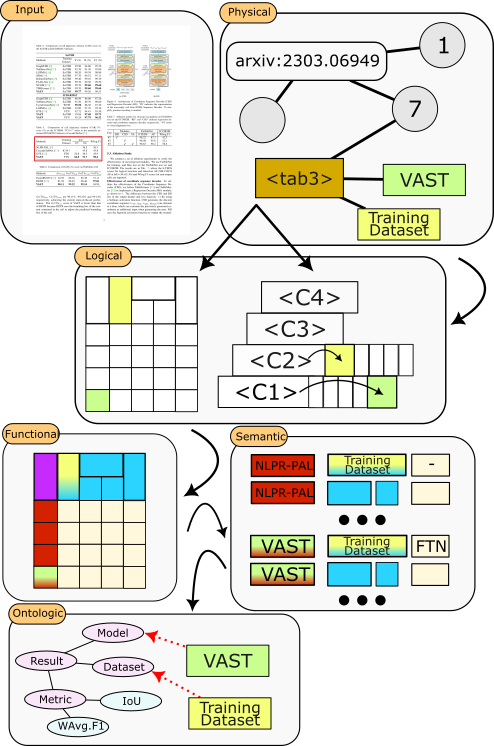}
\caption{Representation of a table extraction workflow using \textit{S2Doc}, showing how each processing stage is modeled across abstraction levels using the example table also shown in Fig.~\ref{fig:example}.
The physical structure is built based on the input PDF document and the output of a table detection system.
A table structure recognition module builds the logical structure in one of two ways.
The functional structure marks header cells and the semantic structure can be generated based on the previous results.
The ontological structure enables the annotation of elements on every level with knowledge.}
\label{fig:workflow}
\end{figure}

The following example illustrates (see Figure~\ref{fig:workflow}) the workflow of how \textit{S2Doc} can be used throughout a table understanding pipeline.
The first step is to define a \texttt{Document} with an identifier (for example, \texttt{arxiv:2303.06949}) and any additional metadata.
The original source contains 17 pages, so 17 \texttt{Page} objects are created.
The table detection (TD) system identifies the example table on page 7 and creates a \texttt{Table} element.
The assumed pipeline is bottom-up, so the first step is to recognize the cells.
This is easy here because the input medium is a PDF document, and the physical model of the document already contains elements for all text lines.
So for every text line that is located within the table region, a \texttt{TableCell} element is created and associated with the table.

The next step is to recognize the table structure (TSR), meaning ordering the cells into a grid, and thereby creating the logical model of the table.
As mentioned above, this can be achieved in different ways.
In our example, the system groups cells by vertical and horizontal alignment to create \texttt{Row} and \texttt{Column} elements; ambiguous entries (e.g., 'IoU', 'Methods', 'WAvg.F1') are temporarily placed in extra rows or columns.
This results in an intermediate logical model (not shown).
A post-processing step resolves these ambiguities by converting temporary entries into spanning cells.
The final logical model assigns each cell to every row and column to which it belongs.

Functional table analysis (TFA) adds an attribute to the \texttt{data} dictionary of each cell to indicate a row or column label.
For example, if one was unsure if 'Training Dataset' is a column header or not, a confidence score for this attribute could be created.
If more context from the surrounding document was acquired, this could be increased or decreased accordingly.

The semantic model is generated using a deterministic algorithm that iterates over all data cells in the table and retrieves their associated row and column headers.
It has to be noted that things like projected row headers in the middle of the table would require more complex logic to be handled correctly, but it is possible to model this using \textit{S2Doc}.

Finally, elements are annotated with known concepts and entities, forming the ontological model.

\section{Conclusion}

To the best of our knowledge, \textit{S2Doc} is the first approach of its kind that combines spatial, logical, and semantic information in a single format.
While it is already suitable for a wide range of document and table understanding applications, it is designed to be easily extendable to new requirements without breaking existing functionality or compromising compatibility.
Because of this, there should be no need to develop custom datastructures, as \textit{S2Doc} can be adapted to most use cases.
It aims to establish a common ground for modeling documents and tables to simplify implementing, integrating, and evaluating different approaches.

In an ongoing effort, we are working on building a framework and module collection for document and table understanding pipelines using \textit{S2Doc} as the central data structure.
Because of that, we are continuously checking \textit{S2Doc} for shortcomings and potential improvements in real-world applications.
The collection will also include tools to convert between existing formats and \textit{S2Doc} to make adopting \textit{S2Doc} for existing pipelines easier.

\section{Bibliographical References}\label{sec:reference}
\vspace{-1cm}
\bibliographystyle{lrec2026-natbib}
\bibliography{lrec2026-example}

\begin{thebibliography}{7}
\expandafter\ifx\csname natexlab\endcsname\relax\def\natexlab#1{#1}\fi

\bibitem[{Breuel(2007)}]{breuel_hocr_2007}
T.~Breuel. 2007.
\newblock \href {https://doi.org/10.1109/ICDAR.2007.4377078} {The {hOCR} microformat for {OCR} workflow and results}.
\newblock In \emph{Ninth International Conference on Document Analysis and Recognition ({ICDAR} 2007) Vol 2}, pages 1063--1067. {IEEE}.
\newblock {ISSN}: 1520-5363.

\bibitem[{Huang et~al.(2023)Huang, Lu, Chen, Li, Xie, Zhu, Gao, and Peng}]{huang2023improvingtablestructurerecognition}
Yongshuai Huang, Ning Lu, Dapeng Chen, Yibo Li, Zecheng Xie, Shenggao Zhu, Liangcai Gao, and Wei Peng. 2023.
\newblock \href {http://arxiv.org/abs/2303.06949} {Improving table structure recognition with visual-alignment sequential coordinate modeling}.

\bibitem[{Hurst(2000)}]{hurst_interpretation_2000}
Matthew~Francis Hurst. 2000.
\newblock \emph{The Interpretation of Tables in Texts}.
\newblock phdthesis, University of Edinburgh.

\bibitem[{{Library of Congress}(2020)}]{alto}
{Library of Congress}. 2020.
\newblock \href {https://www.loc.gov/standards/alto/} {{ALTO} (analyzed layout and text object)}.

\bibitem[{Lin et~al.(2015)Lin, Maire, Belongie, Bourdev, Girshick, Hays, Perona, Ramanan, Zitnick, and Dollár}]{lin_microsoft_2015}
Tsung-Yi Lin, Michael Maire, Serge Belongie, Lubomir Bourdev, Ross Girshick, James Hays, Pietro Perona, Deva Ramanan, C.~Lawrence Zitnick, and Piotr Dollár. 2015.
\newblock \href {https://doi.org/10.48550/arXiv.1405.0312} {Microsoft {COCO}: Common objects in context}.

\bibitem[{Pletschacher and Antonacopoulos(2010)}]{pletschacher_page_2010}
Stefan Pletschacher and Apostolos Antonacopoulos. 2010.
\newblock \href {https://doi.org/10.1109/ICPR.2010.72} {The {PAGE} (page analysis and ground-truth elements) format framework}.
\newblock In \emph{2010 20th International Conference on Pattern Recognition}, pages 257--260. {IEEE}.

\bibitem[{Wang(1996)}]{wang_tabular_1996}
Xinxin Wang. 1996.
\newblock \emph{Tabular Abstraction, Editing, and Formatting}.
\newblock phdthesis, University of Waterloo.

\end{thebibliography}

\end{document}